\documentclass[a4paper,11pt]{article}
\usepackage{pos}

\title{Sensitivity of a Missing-Mass Search for a Light Dark Photon
with a Positron Beam on a Hydrogen Target}

\author*[a]{W.~Xiong}
\author[b]{A.~Gasparian}
\author[c]{B.~Wojtsekhowski}

\affiliation[a]{Shandong University,\\
  Qingdao, Shandong 266237, China}

\affiliation[b]{North Carolina Agricultural and Technical State University,\\
Greensboro, NC 27411 USA}

\affiliation[c]{Thomas Jefferson National Accelerator Facility,\\
Newport News, VA 23606, USA}

\emailAdd{xiongw@sdu.edu.cn}
\emailAdd{agaspari@ncat.edu}
\emailAdd{bogdanw@jlab.org}

\abstract{We present a Geant4-based simulation study of the sensitivity to a light dark photon, $A'$, using a positron beam incident on a thin liquid-hydrogen target and a high-resolution forward electromagnetic calorimeter for final-state photon reconstruction. 
This study is motivated by ongoing research and development toward a low-energy positron facility at Jefferson Lab. 
The proposed search employs the missing-mass technique in the annihilation-in-flight process $e^+e^- \to \gamma A'$, providing sensitivity to the $A'$ independently of its decay mode. 
We evaluate the expected backgrounds from standard QED annihilation and bremsstrahlung processes and present the projected sensitivity to the kinetic-mixing parameter $\epsilon^2$ as a function of the $A'$ mass.
The experiment with 500~MeV beam will provide the data for $A'$ mass range 7-19 MeV.
The projected sensitivity on $\epsilon^2$ is $4\times10^{-8}$ at $m_{A'}=19$~MeV.}

\FullConference{International Workshop on Low Energy Electron Positron Physics at Jefferson Lab (LEEPP2026)\\
23--27 March 2026\\
Thomas Jefferson National Accelerator Facility, Virginia, USA\\}

\begin{document}
\maketitle

\section{Introduction}
\label{sec:intro}

The nature of dark matter remains one of the central open problems in particle physics and cosmology~\cite{Bertone:2004pz}. The absence of a confirmed detection for weakly interacting massive particles (WIMPs) has motivated broader searches for light dark-sector states that interact feebly with ordinary matter~\cite{Schumann:2019eaa,Battaglieri:2017aum}. One well-motivated possibility is the dark photon, $A'$, a hypothetical massive gauge boson associated with an additional $U(1)$ gauge symmetry~\cite{Fayet:1980ad,Fayet:2007ua}. Through kinetic mixing with the Standard Model photon~\cite{Holdom:1985ag}, the $A'$ acquires a suppressed coupling to the electromagnetic current, characterized by the kinetic-mixing parameter $\epsilon$. Such a mediator has been widely considered in models of light dark matter~\cite{Boehm:2003bt,Arkani-Hamed:2008hhe,Pospelov:2008zw}. The present study focuses on the mass range $m_{A'} = 7$--19~MeV.

Existing constraints on the dark-photon mass $m_{A'}$ and kinetic-mixing
parameter $\epsilon$ have been obtained from precision measurements of the
electron and muon anomalous magnetic moments, rare-meson decays, beam-dump
experiments, and $e^+e^-$ collider searches
~\cite{Bjorken:2009mm,Essig:2010xa,Fabbrichesi:2020wbt}. These
measurements constrain complementary regions of parameter space and
generally depend on the assumed $A'$ production and decay properties.
Searches for visible decays, such as $A'\to e^+e^-$, provide strong
constraints when decays to Standard Model particles dominate. However, if
a lighter dark-sector particle $\chi$ exists and
$m_{A'}>2m_\chi$, the decay $A'\to\chi\bar{\chi}$ becomes kinematically
allowed and may dominate when the dark-sector coupling is sufficiently
large. In this case, the $A'$ decays invisibly and the sensitivity of
conventional visible-decay searches is substantially reduced~\cite{Fabbrichesi:2020wbt}. Intermediate scenarios, in which dark-sector
particles subsequently decay into partially visible final states, can
also evade standard invisible searches because visible decay products may
activate detector vetoes or alter the missing-energy signature~\cite{NA64:2021acr,Mongillo:2023hbs}. A broad experimental program has therefore been developed to explore different dark-photon production and decay scenarios. Visible-decay
searches at fixed-target facilities, including multiple experiments at
Jefferson Lab, as well as DarkLight and MAMI, probe dark photons through
reconstructed $e^+e^-$ final states~\cite{APEX:2011dww,HPS:2018xkw,
Dutta:2023ifr,DarkLight:2022uji,Merkel:2014avp}. Missing-energy
measurements, such as those performed by NA64, provide sensitivity to
invisible and semi-visible dark-sector scenarios~\cite{NA64:2019qap, NA64:2021acr,Mongillo:2023hbs}.
Collider searches have also probed visible and invisible dark-photon and related dark-sector signatures over complementary regions of parameter space, while future asymmetric-collider concepts may extend the accessible mass and coupling ranges further~\cite{BaBar:2014zli,KLOE-2:2016ydq,BESIII:2022oww,Belle-II:2022jyy,Morozov:2022sus}.

A complementary approach is the missing-mass technique applied to
positron annihilation in flight,
$e^+e^-\to\gamma A'$,
on atomic electrons in a fixed target. The missing mass is reconstructed
from the known incident-positron four-momentum and the measured
four-momentum of the final-state photon. Since the $A'$ decay products do
not need to be reconstructed, this method provides sensitivity to
invisible, semi-visible, and visible decay modes, with the signal
identified through the production kinematics rather than the decay
topology. This technique was proposed for a low-energy positron beam at
Jefferson Lab~\cite{Wojtsekhowski:2006report,Wojtsekhowski:2009vz} and
for the internal gas target of the VEPP--3 storage ring. For the latter, a
sensitivity of $\epsilon^2\sim(0.2\text{--}1.0)\times10^{-7}$ was
projected over the approximate mass range $m_{A'}=5$--20~MeV
~\cite{Wojtsekhowski:2012zq,Rachek:2017gdc}. A similar approach has been
implemented by the PADME experiment at the INFN Frascati National
Laboratories, although its bunched positron beam limits the achievable
integrated luminosity and increases pileup and accidental backgrounds
~\cite{PADME-2014,PADME:2022xly}. Building on this concept, a full-scale experiment has been proposed using
the planned CEBAF positron beam~\cite{Accardi:2020swt}, with beam energies
of 2.2--11~GeV, together with an electromagnetic calorimeter based on the
PRad apparatus~\cite{Xiong:2019umf}. The proposed experiment targets the
approximate mass range $m_{A'}=15$--90~MeV and projects sensitivity to
$\epsilon^2$ at the level of $2\times10^{-8}$ in part of this range
~\cite{Achenbach:2024pac52}.

A high-current positron beam suitable for such a fixed-target program is
not yet available at Jefferson Lab, although research and development
toward a high-duty-cycle positron source for CEBAF is underway~\cite{Accardi:2020swt,PEPPo:2016saj}. Motivated by this development, the
present work investigates a lower-energy implementation of the
missing-mass technique. We use a dedicated Geant4-based simulation~\cite{GEANT4:2002zbu} of a 500~MeV positron beam incident on a thin
liquid-hydrogen target, with the final-state photon detected by a
segmented, high-resolution forward electromagnetic calorimeter. The
sensitivity to the process $e^+e^-\to\gamma A'$ is evaluated over the
mass range $m_{A'} = 7$--19~MeV. The principal backgrounds from
ordinary two-photon annihilation and positron bremsstrahlung are
simulated, and realistic calorimeter energy and position resolutions are
included. Two complementary event-selection strategies are considered:
an inclusive selection covering the full forward acceptance and a
restricted angular selection guided by the kinematics of two-photon
annihilation. The resulting missing-mass spectra are used to project the
achievable sensitivity to $\epsilon^2$ as a function of $m_{A'}$ for a
representative running scenario.

\section{Experimental Setup and Simulation}
\label{sec:setup}

\begin{figure}[ht]
\centering
\includegraphics[width=0.75\linewidth]{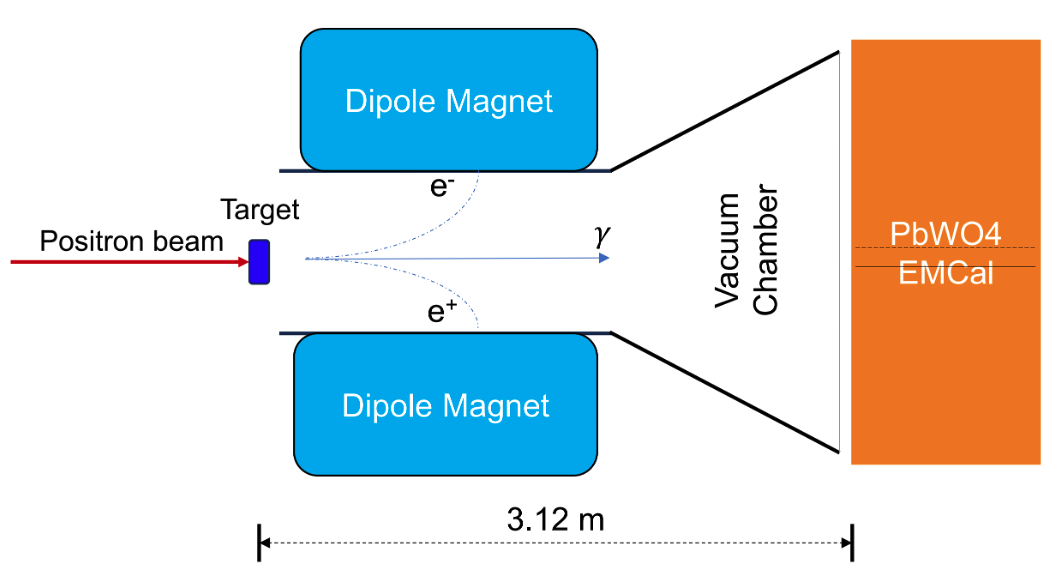}
\caption{Schematic layout of the proposed measurement. The target is a 1~cm long liquid hydrogen, contained in a target cell with 100~$\mu$m Kapton windows. The distance between the target center and front-end of the PbWO$_{4}$ calorimeter is 3.12~m. A sweeper magnet immediately downstream of the target is used to remove the charged particles. A vacuum chamber is connected with the target chamber and extended to the front of the calorimeter.}
\label{fig:apparatus}
\end{figure}

The simulated apparatus follows the fixed-target implementation of the missing-mass concept first proposed in Ref.~\cite{Wojtsekhowski:2006report} and further developed in Ref.~\cite{Wojtsekhowski:2009vz}. A 500~MeV positron beam impinges on a thin (1~cm) hydrogen target with 100~$\mu$m-thick Kapton windows and annihilates with atomic electrons. A dipole magnet installed immediately downstream of the target deflects the unreacted positron beam and other charged particles away from the forward acceptance, while photons produced in the interaction are detected by the high-resolution calorimeter. A vacuum chamber extends through the aperture of the sweeper magnet toward the calorimeter, which minimizes the material traversed by the photons and thereby reduces photon conversions and secondary interactions along the flight path. A schematic layout of the setup is illustrated in Fig.~\ref{fig:apparatus}.

The photon detector is based on the PbWO$_4$ electromagnetic calorimeter previously used in the PRad experiment~\cite{Xiong:2019umf}. The calorimeter consists of 1152 modules, each with approximate dimensions of $2\times2\times18$~cm$^3$. Based on the measured PRad performance, the simulation assumes an energy resolution of $2.6\%/\sqrt{E}$ and a position resolution of $2.5~\mathrm{mm}/\sqrt{E}$, where $E$ is expressed in GeV. The modules are arranged nominally in a $34\times34$ array, corresponding to an overall transverse area of approximately $68\times68$~cm$^2$. The central $2\times2$ modules are omitted, leaving a $4\times4$~cm$^2$ opening for the beam pipe. The 12 modules immediately surrounding the central opening are covered by a tungsten absorber to protect them from excessive radiation. This shielding further reduces the usable photon acceptance near the beam axis.

The calorimeter is positioned 3.12~m downstream of the target. This distance is chosen to provide sufficient angular coverage for the simultaneous detection of both photons from ordinary two-photon annihilation, $e^+e^-\to\gamma\gamma$. The resulting photon polar-angle acceptance begins at approximately $0.7^\circ$--$1.0^\circ$, depending on the azimuthal direction at the boundary of the central shielded region. At the outer boundary of the calorimeter, the acceptance extends to approximately $6.2^\circ$ along the horizontal and vertical edges and to about $8.8^\circ$ at the corners because of the square geometry of the active area.

\section{Analysis Method}
\label{sec:method}

\begin{figure}[ht]
\centering
\includegraphics[width=0.75\linewidth]{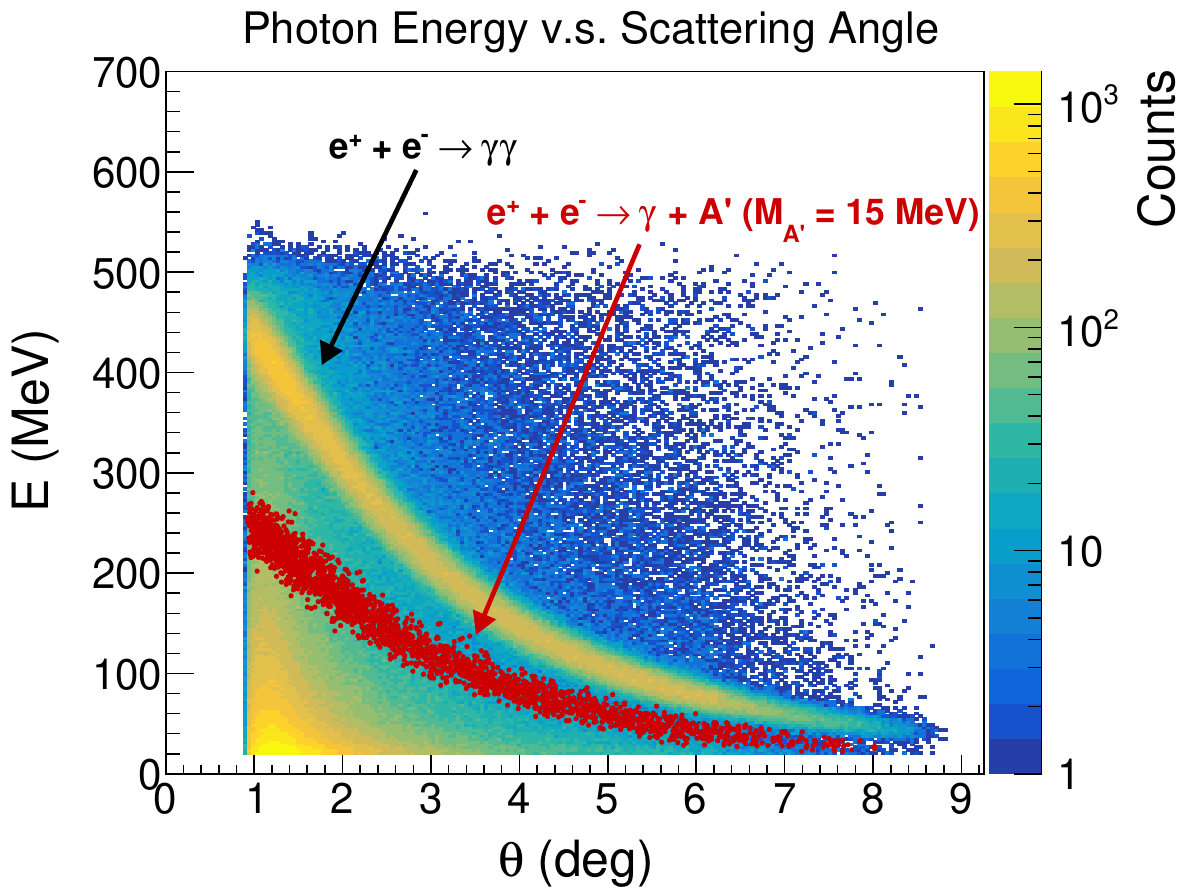}
\caption{Simulated photon energy as a function of laboratory polar angle for reconstructed photon clusters reaching the calorimeter. The dark-photon signal for $m_{A'}=15$~MeV and the benchmark coupling $\epsilon^2=10^{-2}$ is overlaid in red. The dominant two-photon-annihilation background forms a distinct kinematic band, while the $A'$ signal appears at lower photon energy for a given angle.}
\label{fig}
\end{figure}

For fixed-target annihilation of a positron with an atomic electron initially at rest, the missing mass recoiling against the detected photon can be reconstructed from the photon energy and direction. Let $E_+$ and $p_{\rm beam}=\sqrt{E_+^2-m_e^2}$ denote the total energy and momentum of the incident positron, respectively. The total initial-state energy is then $E_{\rm init}=E_++m_e$, and the squared center-of-mass energy is
$s=E_{\rm init}^2-p_{\rm beam}^2$. For a photon with laboratory energy $E_\gamma$ and polar angle $\theta$ relative to the incident beam direction, the missing-mass squared is
\begin{equation}
M_X^2
=s-2E_\gamma
\left(E_{\rm init}-p_{\rm beam}\cos\theta\right).
\label{eq}
\end{equation}
A dark-photon signal would therefore appear as a localized peak at
$M_X^2=m_{A'}^2$ above the continuum background from ordinary QED processes, dominated by two-photon annihilation and positron bremsstrahlung in the target. Figure~\ref{fig} illustrates the corresponding distributions in the $(\theta,E_\gamma)$ plane. The simulated background exhibits a prominent band associated with two-photon annihilation, while the $m_{A'}=15$~MeV signal, generated using the benchmark coupling $\epsilon^2=10^{-2}$, forms a separate band at lower photon energy for a given angle.

Signal events are generated with a dedicated unweighted event generator implementing the differential cross section for $e^+e^-\to\gamma A'$ given in Eq.~(55) of Ref.~\cite{Fayet:2007ua}. Background events are obtained from a simplified Geant4 simulation, including positron bremsstrahlung and both two- and three-photon annihilation. The signal and background samples are then smeared according to the calorimeter energy and position resolutions specified in Sec.~\ref{sec:setup}, ensuring that the same parameterized detector response is applied before event selection.

Two complementary event selections are compared. The \emph{all-photon} selection accepts every reconstructed photon cluster within the usable calorimeter acceptance with energy above a 50~MeV threshold. This method can search for $A'$ regardless of its decay mode. The \emph{single-photon} selection requires exactly one reconstructed cluster with energy above 50~MeV and within a restricted angular range $1.2^\circ<\theta<5.6^\circ$. This interval lies within the inner and outer acceptance boundaries described in Sec.~\ref{sec:method}. For two-photon annihilation at a beam energy of 500~MeV, a photon emitted at $\theta\simeq1.2^\circ$ is accompanied by a second photon at approximately $5.6^\circ$, and vice versa. Thus, throughout this angular interval, both photons from an ideal $e^+e^-\to\gamma\gamma$ event are expected to fall within the active calorimeter acceptance. Requiring exactly one reconstructed cluster therefore vetoes most of the two-photon-annihilation background.

For an invisibly decaying $A'$, the recoil photon is the only detectable final-state particle, so the signal naturally satisfies the single-cluster requirement. The single-photon selection therefore enhances sensitivity to invisible decays by strongly suppressing the two-photon-annihilation background. This improvement comes at the cost of reduced geometric acceptance and diminished sensitivity to visible $A'$ decays, since additional clusters produced by the $A'$ decay products may cause otherwise valid signal events to be rejected. Relative to the all-photon selection, the single-photon selection should therefore be regarded as an analysis optimized primarily for invisible dark-photon decays.

\section{Results}
\label{sec:results}
\begin{figure}[ht]
  \centering
  \begin{minipage}{0.49\linewidth}
    \centering
    \includegraphics[width=\linewidth]{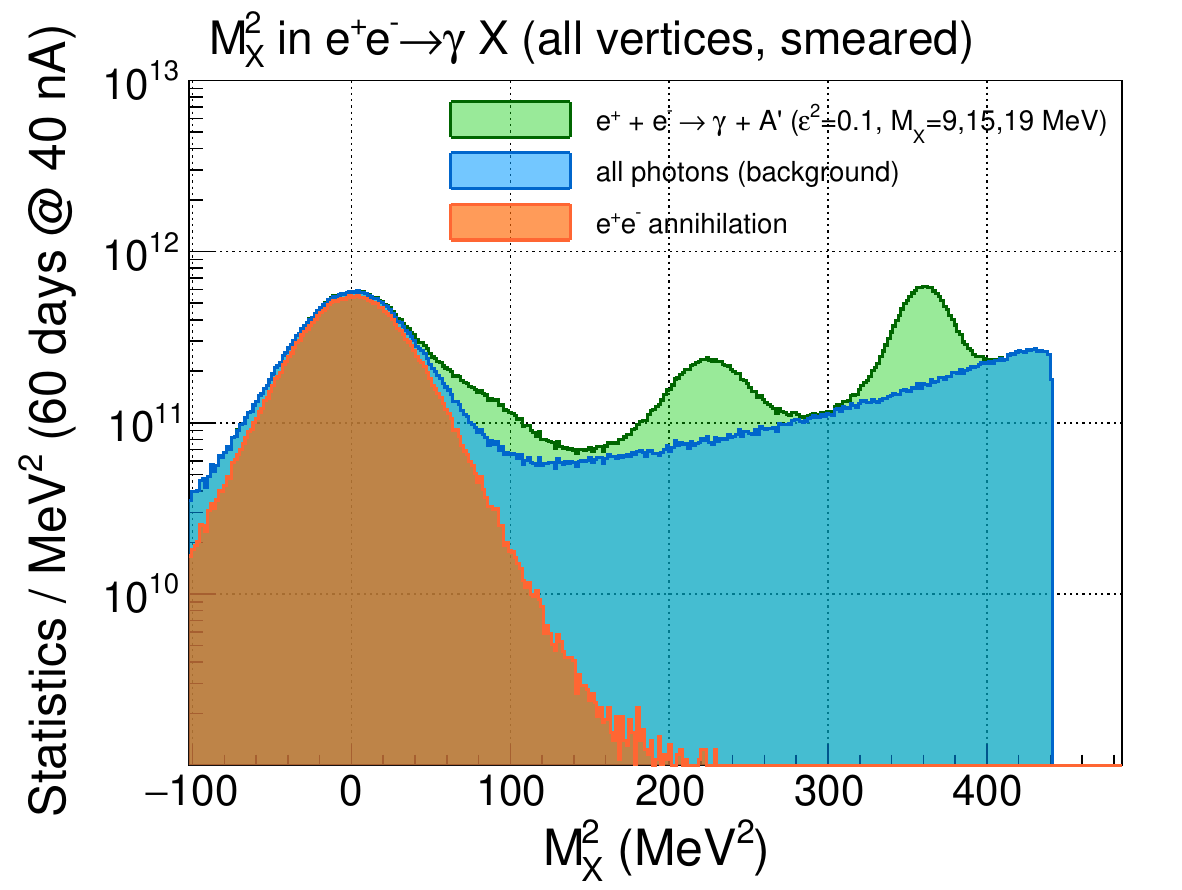}\\
    \small (a) all-photon selection
  \end{minipage}\hfill
  \begin{minipage}{0.49\linewidth}
    \centering
    \includegraphics[width=\linewidth]{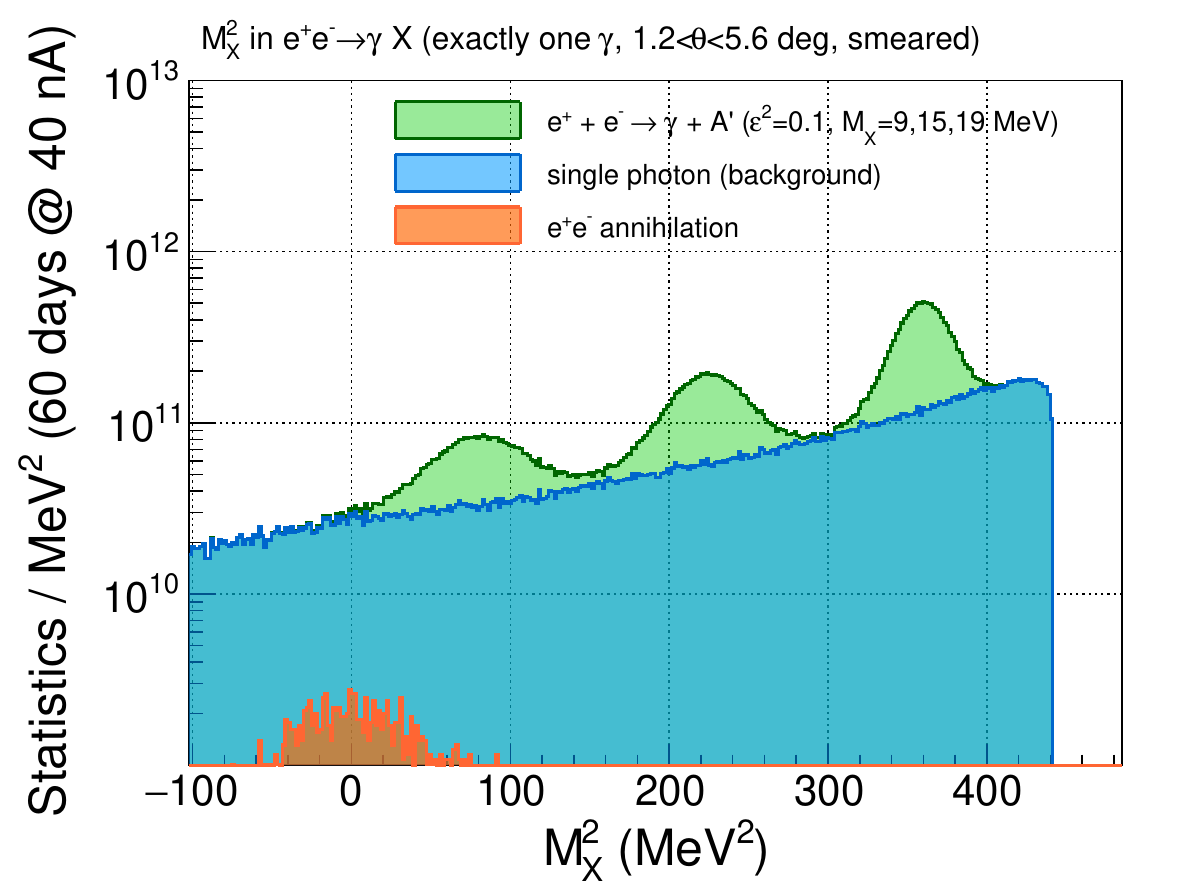}\\
    \small (b) single-photon selection
  \end{minipage}
  \caption{Reconstructed $M_X^2$ spectra for (a) the all-photon selection and (b) the single-photon selection, assuming a 60-day run with a 40~nA positron beam. The distributions show all selected photon clusters (blue), the positron-annihilation contribution (orange), and the spectra including dark-photon signals with $m_{A'}=9$, 15, and 19~MeV generated at $\epsilon^2=0.1$ (green). The signal coupling is enlarged for visibility. The single-photon requirement strongly suppresses the two-photon-annihilation peak near $M_X^2=0$.}
  \label{fig:mx2spectra}
\end{figure}
For the sensitivity projection, we assume a 500~MeV, 40~nA positron beam incident on the 1~cm-long liquid-hydrogen target described in Sec.~\ref{sec:setup}, with a total running time of 60~days. Figure~\ref{fig:mx2spectra}(a) shows the reconstructed $M_X^2$ spectrum obtained with the all-photon selection. Dark-photon signals with $m_{A'}=9$, 15, and 19~MeV are overlaid using $\epsilon^2=0.1$ for visibility; this value is much larger than the projected sensitivity. The spectrum is dominated by the two-photon-annihilation peak near $M_X^2=0$, together with a broader continuum arising primarily from positron bremsstrahlung and three-photon annihilation. Consequently, the all-photon selection has reduced sensitivity to low-mass signals located near the two-photon-annihilation peak. Figure~\ref{fig:mx2spectra}b shows the corresponding spectrum for the single-photon selection. Requiring exactly one reconstructed cluster within the angular range $1.2^\circ<\theta<5.6^\circ$ removes most of the two-photon-annihilation background as discussed in Sec.~\ref{sec:method}. The peak near $M_X^2=0$ is suppressed by approximately two orders of magnitude relative to the all-photon selection, making low-mass signals substantially more prominent above the remaining background.

\begin{figure}[ht]
  \centering
  \begin{minipage}{0.49\linewidth}
    \centering
    \includegraphics[width=\linewidth]{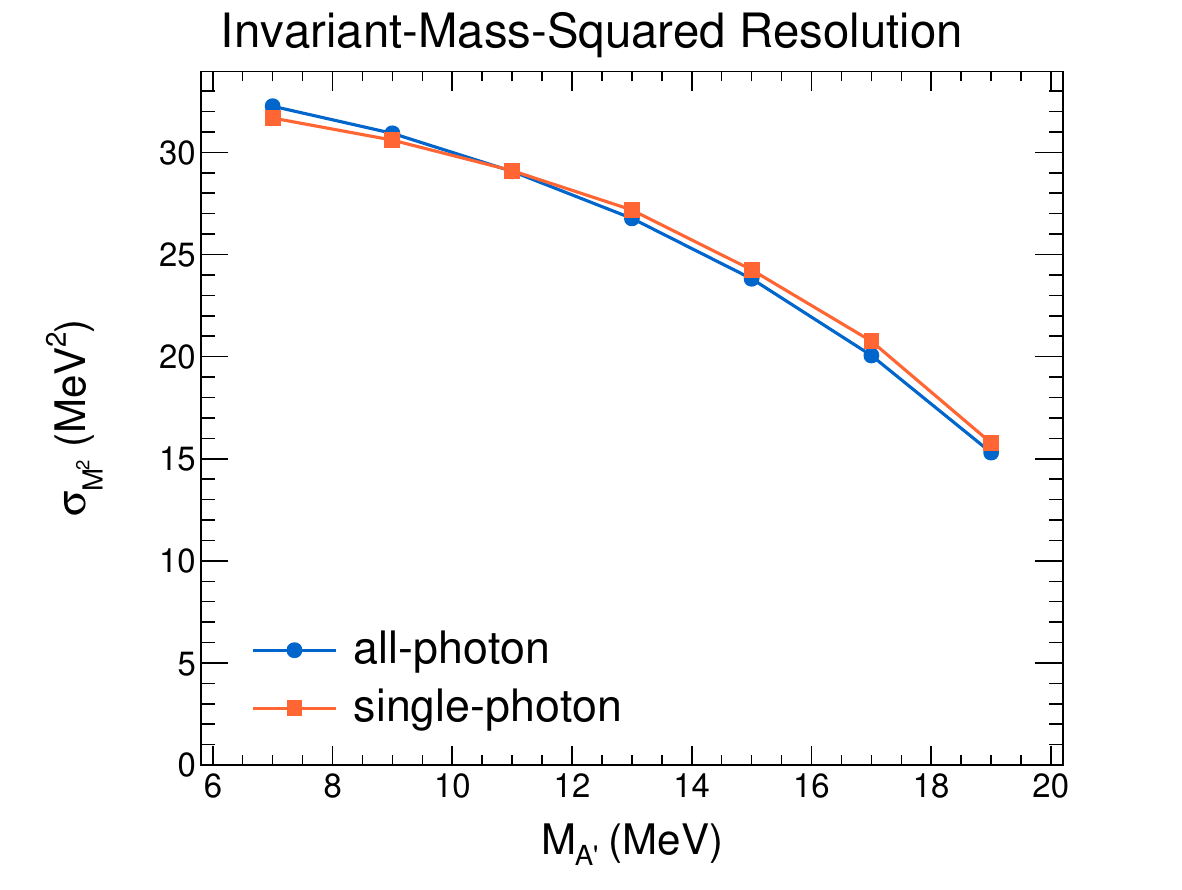}\\
    \small (a) $M_X^2$ resolution
  \end{minipage}\hfill
  \begin{minipage}{0.49\linewidth}
    \centering
    \includegraphics[width=\linewidth]{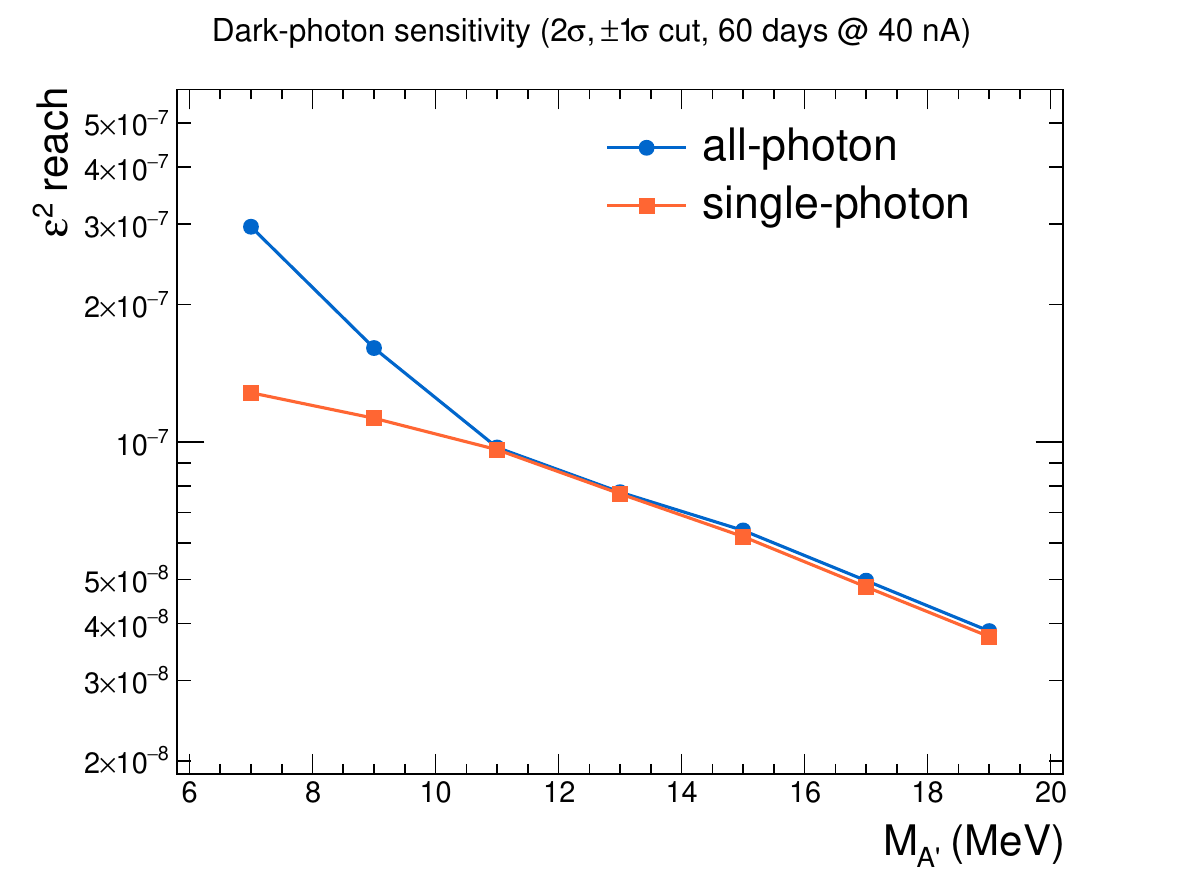}\\
    \small (b) $\epsilon^2$ sensitivity
  \end{minipage}
  \caption{(a) Fitted missing-mass-squared resolution, $\sigma_{M^2}$, as a function of $m_{A'}$, and (b) projected $\epsilon^2$ sensitivity based on $S/\sqrt{B}=2$, assuming a 60-day run with a 40~nA positron beam. Results are shown for the all-photon and single-photon selections.}
  \label{fig:results2}
\end{figure}

Figure~\ref{fig:results2}a shows the fitted missing-mass-squared resolution, $\sigma_{M^2}$, as a function of $m_{A'}$ for the two selections. The resolution is extracted from a Gaussian fit to the simulated signal peak. The two selections give nearly identical resolutions, which improve from approximately 32~MeV$^2$ at $m_{A'}=7$~MeV to approximately 15~MeV$^2$ at $m_{A'}=19$~MeV.

To estimate the statistical sensitivity, a
$\pm1\sigma_{M^2}$ signal window is defined around each fitted peak and count the generated signal $N_{S_0}$ (generated with coupling constant $\epsilon_0$) and background $N_B$ in that window. Since the signal yield scales as $N_S\propto\epsilon^2$, the coupling corresponding to an approximate statistical significance $N_S/\sqrt{N_B}=Z$ is
\begin{equation}
\epsilon^2
= \epsilon_0^2 Z\frac{\sqrt{N_B}}{N_{S_0}}.
\label{eq:sens}
\end{equation}

Figure~\ref{fig:results2}(b) shows the projected sensitivity obtained from Eq.~\eqref{eq:sens} using $Z=2$. At low masses, the single-photon selection improves the sensitivity by more than a factor of two. For example, at $m_{A'}=7$~MeV, the projected reach improves from approximately $\epsilon^2=3\times10^{-7}$ for the all-photon selection to approximately $1.3\times10^{-7}$ for the single-photon selection. This improvement results from the strong suppression of the two-photon-annihilation background near $M_X^2=0$, and could be extended to even lower $A'$ mass.

Above $m_{A'}\approx12$~MeV, the sensitivities obtained with the two selections become comparable, reaching approximately $\epsilon^2=4\times10^{-8}$ at $m_{A'}=19$~MeV. In this higher-mass region, the signal peaks are sufficiently separated from the two-photon-annihilation peak that this background is already small, and the additional single-cluster veto provides little further improvement. The enhanced low-mass reach of the single-photon selection therefore applies primarily to invisible $A'$ decays. For visible decays, additional calorimeter activity may cause signal events to fail the single-cluster requirement, making the all-photon selection the more inclusive baseline.


\section{Conclusions}
\label{sec:conclusions}

We have presented a Geant4-based simulation study of a missing-mass search for a light dark photon, $A'$, through the annihilation-in-flight process $e^+e^-\to\gamma A'$. The proposed setup uses a 500~MeV positron beam incident on a 1~cm-long liquid-hydrogen target, with the recoil photon detected by the PRad PbWO$_4$ calorimeter positioned 3.12~m downstream.

Two complementary analysis strategies were investigated. The inclusive all-photon selection retains broad sensitivity to different $A'$ decay scenarios, whereas the single-photon selection suppresses the dominant two-photon-annihilation background and is optimized primarily for invisible $A'$ decays. The single-photon requirement improves the projected sensitivity to $\epsilon^2$ by more than a factor of two at the lowest mass considered, $m_{A'}\approx7$~MeV. Note that with the prominent annihilation peak removed, it is possible to extend the search to even lower $A'$ mass. 
Above $m_{A'}\approx12$~MeV, the sensitivities of the two selections become comparable because the signal is sufficiently separated from the two-photon-annihilation peak. 
For a representative 60-day run with a 40~nA positron beam, the projected sensitivity ranges from approximately $\epsilon^2=1.3\times10^{-7}$ at $m_{A'}=7$~MeV to approximately $4\times10^{-8}$ at $m_{A'}=19$~MeV.
This is 10-100 times more sensitive than the limits obtained from the studies of the magnetic moment ($g-2$) of an electron and muon. 

These results demonstrate the feasibility of a low-energy positron-beam search for light dark photons at Jefferson Lab and illustrate the effectiveness of kinematic and cluster-based selections for suppressing the dominant QED backgrounds. The same missing-mass concept, implemented with the proposed higher-energy CEBAF positron beam and a PRad-based calorimeter, forms the basis of the PAC52 proposal to extend the accessible dark-photon mass range up to approximately 90~MeV~\cite{Achenbach:2024pac52}. 
Realization of such a program will depend on the development of a high-duty-cycle positron source for CEBAF~\cite{Accardi:2020swt,PEPPo:2016saj}.

\section*{Acknowledgments}
We appreciate the efforts of the positron working group, lead by D.~Higinbotham, J.~Grames, and E.~Voutier, 
for their efforts in the development of the c.w. positron beam at JLab, and also D.~Mack for fruitful discussion about the signal rate. 
W.X. is supported by the Shandong Province Natural Science Foundation under Grant No.~2023HWYQ-010,
A.G. is supported by the U. S. National Science Foundation (NSF MRI PHY-1229153),
B.W. is supported by the U.S. Department of Energy, Office of Science, Office of Nuclear Physics under Contract No. 89243126CSC000213.



\bibliographystyle{unsrt}
\bibliography{references}

\end{document}